\documentclass[graybox]{svmult}
\usepackage{mathptmx}
\usepackage{helvet}
\usepackage{courier}
\usepackage[latin9]{inputenc}
\usepackage{color}
\usepackage{mathrsfs}
\usepackage{url}
\usepackage{amsmath}
\usepackage{amssymb}
\usepackage{graphicx}
\usepackage{esint}
\usepackage{microtype}
\usepackage[unicode=true,
 bookmarks=false,
 breaklinks=false,pdfborder={0 0 1},backref=none,colorlinks=true]
 {hyperref}
\hypersetup{
 allcolors=blue}

\makeatletter
\usepackage{type1cm}
\usepackage{makeidx}
\usepackage{multicol}
\usepackage[bottom]{footmisc}
\usepackage{amsfonts}\usepackage{color}
\usepackage{tikz}

\usepackage{mathrsfs}
\DeclareMathAlphabet{\mathcal}{OMS}{cmsy}{m}{n}

\newcommand{\electrostaticPotential}{\Phi}

\makeatother

\begin{document}
\title*{Non-isothermal Scharfetter--Gummel scheme\\
for electro-thermal transport simulation in degenerate semiconductors}
\titlerunning{Non-isothermal Scharfetter-{}-Gummel scheme for degenerate semiconductors}
\author{Markus Kantner and Thomas Koprucki}
\institute{Markus~Kantner\;$\cdot$\;Thomas~Koprucki \at Weierstrass Institute
(WIAS), Mohrenstr. 39, 10117 Berlin, Germany\\
 e-mail: \href{mailto:kantner\%40wias-berlin.de}{kantner@wias-berlin.de},
\href{mailto:koprucki\%40wias-berlin.de}{koprucki@wias-berlin.de}}
\maketitle

\abstract{Electro-thermal transport phenomena in semiconductors are described
by the non-isothermal drift-diffusion system. The equations take a
remarkably simple form when assuming the Kelvin formula for the thermopower.
We present a novel, non-isothermal generalization of the Scharfetter--Gummel
finite volume discretization for degenerate semiconductors obeying
Fermi--Dirac statistics, which preserves numerous structural properties
of the continuous model on the discrete level. The approach is demonstrated
by 2D simulations of a heterojunction bipolar transistor. }

\keywords{Scharfetter--Gummel scheme\,$\cdot$\,Fermi--Dirac statistics\,$\cdot$\,electro-thermal
transport\,$\cdot$\,non-isothermal drift-diffusion system\,$\cdot$\,Seebeck
effect\,$\cdot$\,self-heating \\[5pt] \textbf{MSC }(2010)\textbf{:}
35K05\,$\cdot$\,35K08\,$\cdot$\,35Q79\,$\cdot$\,65N08\,$\cdot$\,80M12\,$\cdot$\,82B35\,$\cdot$\,82D37 }



\section{Introduction}

Self-heating effects are a major concern in modern semiconductor devices,
where the on-going miniaturization of feature size leads to increased
power loss densities. The optimal design of semiconductor devices
relies on numerical simulations, based on thermodynamically consistent
models for the coupled electro-thermal transport processes. The standard
model for the simulation of self-consistent charge and heat transport
processes is the non-isothermal drift-diffusion system \cite{Albinus2002,Kantner2020a,Wachutka1990},
which couples the semiconductor device equations to a heat transport
equation. The magnitude of the thermoelectric cross effects (Seebeck
effect, Thomson--Peltier effect) is governed by the Seebeck coefficient
(also \emph{thermopower}), which quantifies the thermoelectric voltage
induced by a temperature gradient. Recently \cite{Kantner2020a},
the non-isothermal drift-diffusion system has been studied assuming
the so-called \emph{Kelvin formula} for the thermopower \cite{Peterson2010},
which has two important implications: First, the Seebeck term in the
current density expressions can be entirely absorbed in a temperature-dependent
diffusion constant via a generalized Einstein relation. Second, the
heat generation rate involves solely the three classically known self-heating
effects without any further (transient) contribution. The model equations
and its key features are described in Sect.~\ref{sec:Non-isothermal-drift-diffusion-system}.
In Sect.~\ref{sec:Finite-volume-discretization}, we present a finite
volume discretization based on a novel, non-isothermal generalization
of the Scharfetter--Gummel scheme for the discrete fluxes. The scheme
holds for Fermi--Dirac statistics and preserves numerous structural
and thermodynamic properties of the continuous system.

\vspace{-1.5ex}

\section{Non-isothermal drift-diffusion system \label{sec:Non-isothermal-drift-diffusion-system}}

We consider the non-isothermal drift-diffusion system on $\Omega\subset\mathbb{R}^{d}$,
$d\in\left\{ 1,2,3\right\} $, 
\begin{align}
-\nabla\cdot\varepsilon\nabla\Phi & =q\left(C+p-n\right),\label{eq: Poisson equation}\\
q\partial_{t}n-\nabla\cdot\mathbf{j}_{n} & =-qR,\label{eq: electron continuity equation}\\
q\partial_{t}p+\nabla\cdot\mathbf{j}_{p} & =-qR,\label{eq: hole continuity equation}\\
c_{V}\partial_{t}T-\nabla\cdot\kappa\nabla T & =H.\label{eq: heat equation}
\end{align}
Poisson's Eq.~(\ref{eq: Poisson equation}) describes the electrostatic
potential $\Phi$ generated by the electron density $n$, the density
of valence band holes $p$ and the built-in doping profile $C$. Here,
$q$ is the elementary charge and $\varepsilon$ is the (absolute)
permittivity of the material. The transport and recombination dynamics
of the electrons and holes are modeled by the continuity Eqs.~(\ref{eq: electron continuity equation})--(\ref{eq: hole continuity equation}),
where $\mathbf{j}_{n/p}$ are the electrical current densities and
$R$ is the $\text{{(net-)}}$recombination rate, which comprises
several radiative and non-radiative processes \cite{Farrell2017,Palankovski2004}.
The temperature distribution in the device is described by the heat
equation \eqref{eq: heat equation}, where $c_{V}$ is the volumetric
heat capacity, $\kappa$ is the thermal conductivity and $H$ is the
heat generation rate.

The carrier densities are related with the quasi-Fermi potentials
$\varphi_{n/p}$, the electrostatic potential $\Phi$ and the (absolute)
temperature $T$ via the state equations 
\begin{align}
n & =N_{c}\left(T\right)\mathscr{F}\left(\frac{q(\Phi{-}\varphi_{n}){-}E_{c}(T)}{k_{B}T}\right), & p & =N_{v}\left(T\right)\mathscr{F}\left(\frac{E_{v}(T){-}q(\Phi{-}\varphi_{p})}{k_{B}T}\right),\label{eq: state equations}
\end{align}
where $N_{c/v}$ are the effective density of states, $E_{c/v}$ are
the band edge energies of the conduction and the valence band, respectively,
and $k_{B}$ is Boltzmann's constant. The function $\mathscr{F}$
describes the occupation probability of the electronic states. In
the case of non-degenerate semiconductors (Maxwell--Boltzmann statistics),
$\mathscr{F}\left(\eta\right)=\exp{\left(\eta\right)}$ is an exponential
function. At high carrier densities, where degeneration effects due
to the Pauli exclusion principle (Fermi--Dirac statistics) must be
taken into account, $\mathscr{F}$ is typically given by the Fermi--Dirac
integral $F_{1/2}$ \cite{Farrell2017}. The approach outlined below,
does not rely on the specific form of $\mathscr{F}$ and is applicable
to materials with arbitrary density of states and degenerate or non-degenerate
statistics \cite{Kantner2020a}.

\subsection{Kelvin formula for the thermopower\label{sec:Kelvin-formula}}

The electrical current densities are modeled as 
\begin{align}
\mathbf{j}_{n} & =-\sigma_{n}\left(\nabla\varphi_{n}+P_{n}\nabla T\right), & \mathbf{j}_{p} & =-\sigma_{p}\left(\nabla\varphi_{p}+P_{p}\nabla T\right),\label{eq: current densities - thermodynamic form}
\end{align}
where $\sigma_{n/p}$ are the electrical conductivities and $P_{n/p}$
are the thermopowers of the material. In this paper, we choose the
thermopowers according to the Kelvin formula as variational derivatives
of the entropy $\mathcal{S}$ with respect to the carrier densities
\begin{align}
qP_{n} & =-\mathrm{D}_{n}\mathcal{S}\left(n,p,T\right), & qP_{p} & =+\mathrm{D}_{p}\mathcal{S}\left(n,p,T\right),\label{eq: Kelvin formula}
\end{align}
where $\mathrm{D}$ denotes the Gâteaux derivative. The Kelvin formula
is the low frequency and long wavelength limit of the microscopically
exact Kubo formula \cite{Peterson2010}. It was shown to provide a
good approximation for several materials at sufficiently high temperature.
The entropy is obtained from the free energy $\mathcal{F}\left(n,p,T\right)$
of the system.

We assume the free energy functional \cite{Albinus2002,Kantner2020a}
\begin{align}
\mathcal{F}\left(n,p,T\right) & =\int_{\Omega}\mathrm{d}V\,\bigg(k_{B}T\mathscr{F}^{-1}\left(\frac{n}{N_{c}}\right)n-k_{B}TN_{c}\mathscr{G}\left(\mathscr{F}^{-1}\left(\frac{n}{N_{c}}\right)\right)+E_{c}(T)n\label{eq: free energy}\\
 & \hphantom{=\int_{\Omega}\mathrm{d}V\,\bigg(}+k_{B}T\mathscr{F}^{-1}\left(\frac{p}{N_{v}}\right)p-k_{B}TN_{v}\mathscr{G}\left(\mathscr{F}^{-1}\left(\frac{p}{N_{v}}\right)\right)-E_{v}(T)p\bigg)\nonumber \\
 & \hphantom{=}+\int_{\Omega}\mathrm{d}V\,f_{L}\left(T\right)+\frac{1}{2}\int_{\Omega}\mathrm{d}V\int_{\Omega}\mathrm{d}V^{\prime}\,G\left(\mathbf{r},\mathbf{r}^{\prime}\right)\rho\left(\mathbf{r}\right)\rho\left(\mathbf{r}^{\prime}\right)+\int_{\Omega}\mathrm{d}V\,\Phi_{\text{ext}}\rho,\nonumber 
\end{align}
where the first to lines describe the free energy of the non-interacting
electron-hole plasma (quasi-free Fermi gas), $f_{L}$ is the free
energy of the lattice phonons (ideal Bose gas), $\mathscr{G}$ is
the antiderivative of $\mathscr{F}$ (i.e., $\mathscr{G}^{\prime}\left(\eta\right)=\mathscr{F}\left(\eta\right)$),
$G\left(\mathbf{r},\mathbf{r}^{\prime}\right)$ is the Green's function
of Poisson's equation and $\rho=q\left(p-n\right)$ is the mobile
charge density. The potential $\Phi_{\text{ext}}$ is generated by
the built-in doping-profile and the applied bias. 
\begin{figure}[t]
\sidecaption\hfill{}\includegraphics{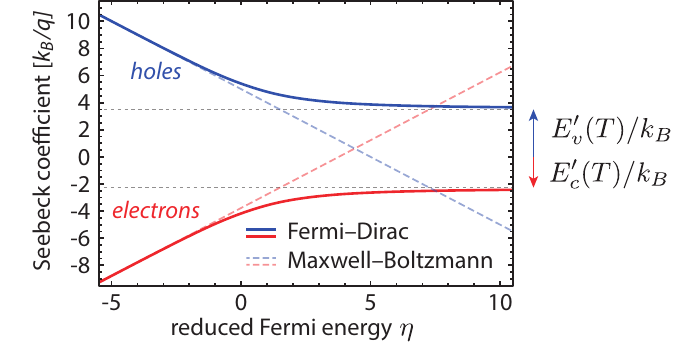}\caption{Thermopowers $P_{n/p}$ according to Eqs.~(\ref{eq: Seebeck coefficients - explicit})
as functions of the reduced Fermi energy $\eta$ (argument of $\mathscr{F}$
in Eqs.~(\ref{eq: state equations})) in units of $k_{B}/q$. The
thermopowers are plotted for $\mathscr{F}\left(\eta\right)=F_{1/2}\left(\eta\right)$
and $N_{c/v}\propto T^{3/2}$. Adapted, with permission, from \cite{Kantner2020a}.}
\label{fig: Kelvin formula} 
\end{figure}

The free energy (\ref{eq: free energy}) recovers the state equations
(\ref{eq: state equations}) via the variational derivative with respect
to the carrier densities $\mathrm{D}_{n/p}\mathcal{F}:=\mp q\varphi_{n/p}$,
which is the defining relation for the quasi-Fermi potentials, see
\cite{Kantner2020a}. The entropy functional is defined as the derivative
of the free energy (\ref{eq: free energy}) with respect to the temperature:
$\mathcal{S}\left(n,p,T\right)=-\partial_{T}\mathcal{F}\left(n,p,T\right).$
Evaluation of Eq.~(\ref{eq: Kelvin formula}) yields the thermopowers
\begin{subequations}\label{eq: Seebeck coefficients - explicit}
\begin{align}
P_{n} & \left(n,T\right)=-\frac{k_{B}}{q}\left(\frac{TN_{c}^{\prime}\left(T\right)}{N_{c}\left(T\right)}g\left(\frac{n}{N_{c}\left(T\right)}\right)-\mathscr{F}^{-1}\left(\frac{n}{N_{c}\left(T\right)}\right)-\frac{1}{k_{B}}E_{c}^{\prime}\left(T\right)\right),\label{eq: Seebeck coefficient - electrons}\\
P_{p} & \left(p,T\right)=+\frac{k_{B}}{q}\left(\frac{TN_{v}^{\prime}\left(T\right)}{N_{v}\left(T\right)}g\left(\frac{p}{N_{v}\left(T\right)}\right)-\mathscr{F}^{-1}\left(\frac{p}{N_{v}\left(T\right)}\right)+\frac{1}{k_{B}}E_{v}^{\prime}\left(T\right)\right).\label{eq: Seebeck coefficient - holes}
\end{align}
\end{subequations} The temperature-dependency of the band edge energies
can be modeled using, e.g., the Varshni model \cite{Kantner2020a,Palankovski2004}.
The function 
\begin{equation}
g\left(x\right)=x\,\left(\mathscr{F}^{-1}\right)^{\prime}\left(x\right)\label{eq: degeneration factor}
\end{equation}
quantifies the degeneration of the carriers ($g>1$ for Fermi--Dirac
statistics; $g\equiv1$ for Maxwell--Boltzmann statistics). See Fig.~\ref{fig: Kelvin formula}
for a plot of the Seebeck coefficients (\ref{eq: Seebeck coefficients - explicit}).

\subsection{Drift-diffusion currents and heat generation rate \label{sec: Drift-diffusion flux and self-heating kernel}}

The Kelvin formula has two important implications, which lead to a
very simple and appealing form of the thermoelectric cross effects
in the system (\ref{eq: Poisson equation})--(\ref{eq: heat equation}).

First, we rewrite the electrical current densities by passing from
the thermodynamic form (\ref{eq: current densities - thermodynamic form})
to the drift-diffusion form. By explicitly evaluating the gradient
of the quasi-Fermi potentials using the state equations (\ref{eq: state equations}),
one observes that the Seebeck terms $\mathbf{j}_{n/p}\vert_{\text{Seebeck}}=-\sigma_{n/p}P_{n/p}\nabla T$
cancel out \emph{exactly} from the expressions \cite{Kantner2020a}.
Using the conductivities $\sigma_{n}=qM_{n}n$ and $\sigma_{p}=qM_{p}p$
(with mobilities $M_{n/p}$), one arrives at 
\begin{align}
\mathbf{j}_{n} & =-qM_{n}n\nabla\Phi+qD_{n}\left(n,T\right)\nabla n, & \mathbf{j}_{p} & =-qM_{p}p\nabla\Phi-qD_{p}\left(p,T\right)\nabla p.\label{eq: drift-diffusion current density}
\end{align}
We emphasize that in Eq.~(\ref{eq: drift-diffusion current density})
-- even though there is no explicit thermal driving force $\propto\nabla T$
-- the Seebeck effect is fully taken into account via the (temperature-dependent)
diffusion coefficients $D_{n/p}$. The latter obey the generalized
Einstein relations \cite{Koprucki2015} 
\begin{align}
qD_{n} & =k_{B}TM_{n}g\left(n/N_{c}\left(T\right)\right), & qD_{p} & =k_{B}TM_{p}g\left(p/N_{v}\left(T\right)\right).\label{eq: generalized Einstein relation}
\end{align}
The flux discretization described in Sect.~\ref{subsec:Generalized-Scharfetter=00003D002013Gummel-scheme}
is based on the drift-diffusion form (\ref{eq: drift-diffusion current density}).

The second implication of the Kelvin formula concerns the heat generation
rate $H$. The commonly accepted model for $H$, which was derived
by Wachutka \cite{Wachutka1990} from linear irreversible thermodynamics,
takes a particularly simple form, when assuming the Kelvin formula
for the thermopower. One obtains (see Appendix)
\begin{equation}
H=\sum_{\lambda\in\left\{ n,p\right\} }\frac{1}{\sigma_{\lambda}}\left\Vert \mathbf{j}_{\lambda}\right\Vert ^{2}-\sum_{\lambda\in\left\{ n,p\right\} }T\,\mathbf{j}_{\lambda}\cdot\nabla P_{\lambda}+q\left(\varphi_{p}+TP_{p}-\varphi_{n}-TP_{n}\right)R,\label{eq: heat source term}
\end{equation}
which involves solely the three classically known self-heating effects,
namely Joule heating (first term), the Thomson--Peltier effect (second
term) and recombination heating (last term). Any further (transient)
contributions, which necessarily arise for thermopowers different
from the Kelvin formula (\ref{eq: Kelvin formula}), do not occur
in the model.

\section{Finite volume discretization\label{sec:Finite-volume-discretization}}

We assume a boundary conforming Delaunay triangulation of the computational
domain $\Omega\subset\mathbb{R}^{d}$, $d=\left\{ 1,2,3\right\} $,
and obtain the finite volume discretization \cite{Farrell2017} of
the (stationary) system (\ref{eq: Poisson equation})--(\ref{eq: heat equation})
by integration over the (restricted) Voronoï cells as \begin{subequations}\label{eq: discrete system}
\begin{align}
-\sum_{L\in N\left(K\right)}s_{K,L}\varepsilon\left(\Phi_{L}-\Phi_{K}\right) & =q\vert\Omega_{K}\vert\left(C_{K}+p_{K}-n_{K}\right),\label{eq: discrete Poisson equation}\\
-\sum_{L\in N\left(K\right)}s_{K,L}J_{n,K,L} & =-q\vert\Omega_{K}\vert R_{K},\label{eq: discrete electron continuity equation}\\
+\sum_{L\in N\left(K\right)}s_{K,L}J_{p,K,L} & =-q\vert\Omega_{K}\vert R_{K},\label{eq: discrete hole continuity equation}\\
-\sum_{L\in N\left(K\right)}s_{K,L}\kappa_{K,L}\left(T_{L}-T_{K}\right) & =\frac{1}{2}\sum_{L\in N\left(K\right)}s_{K,L}\left(H_{J,K,L}+H_{\text{T--P},K,L}\right)+\vert\Omega_{K}\vert H_{R,K}.\label{eq: discrete heat equation}
\end{align}
\end{subequations} Here, $\left|\Omega_{K}\right|$ is the volume
of the $K$-th Voronoï cell, $s_{K,L}=\vert\partial\Omega_{K}\cap\partial\Omega_{L}\vert/\left\Vert \mathbf{r}_{L}-\mathbf{r}_{K}\right\Vert $
is a geometric factor and $N\left(K\right)$ is the set of adjacent
nodes of $K$. The subscripts $K$, $L$ indicate evaluation on the
respective nodes or edges. The discrete heat sources are \begin{subequations}\label{eq: discrete heat source}
\begin{align}
H_{J,K,L} & =-\sum_{\lambda\in\left\{ n,p\right\} }J_{\lambda,K,L}\left(\varphi_{\lambda,L}-\varphi_{\lambda,K}+P_{\lambda,K,L}\left(T_{L}-T_{K}\right)\right),\label{eq: discrete Joule heating}\\
H_{\text{T--P},K,L} & =-\sum_{\lambda\in\left\{ n,p\right\} }T_{K,L}J_{\lambda,K,L}\left(P_{\lambda,L}-P_{\lambda,K}\right),\label{eq: discrete Thomson Peltier heating}\\
H_{R,K} & =q\left(\varphi_{p,K}+T_{K}P_{p,K}-\varphi_{n,K}-T_{K}P_{n,K}\right)R_{K},\label{eq: discrete recombination heating}
\end{align}
\end{subequations} where we used a technique involving a weakly converging
gradient developed in \cite{Eymard2003} for the discretization of
the Joule and Thomson--Peltier terms (see \cite{Kantner2020a} for
details).

\vspace{-3.7ex}

\subsection{Generalized Scharfetter--Gummel scheme \label{subsec:Generalized-Scharfetter=00003D002013Gummel-scheme}}

A robust discretization of the flux projections $J_{n/p,K,L}=\left(\mathbf{r}_{L}-\mathbf{r}_{K}\right)\cdot\mathbf{j}_{n/p}$
is obtained by integrating Eq.~(\ref{eq: drift-diffusion current density})
along the edge $\overline{KL}:=\left\{ \mathbf{r}\left(x\right)=x\,\mathbf{r}_{L}+\left(1-x\right)\,\mathbf{r}_{K},\,x\in\left[0,1\right]\right\} $,
while assuming the electric field, the current density and the mobility
to be constant along $\overline{KL}$. The temperature is assumed
to be an affine function between adjacent nodes: $T\left(x\right)=x\,T_{L}+\left(1-x\right)\,T_{K}$,
$x\in\left[0,1\right]$. In the case of Fermi--Dirac statistics (with
$g\neq1$), the resulting two-point boundary value problem on $x\in\left[0,1\right]$
\cite{Kantner2020a} 
\begin{align*}
k_{B}T(x)g\bigg(\frac{n(x)}{N_{c}\left(T(x)\right)}\bigg)\frac{\mathrm{d}n}{\mathrm{d}x} & =q\left(\Phi_{L}-\Phi_{K}\right)n(x)+\frac{J_{n,K,L}}{M_{n,K,L}}, & n(0) & =n_{K}, & n(1) & =n_{L},
\end{align*}
can be solved approximately, by freezing the degeneracy factor (\ref{eq: degeneration factor})
to a suitable average $g_{n/p,K,L}$ \cite{Bessemoulin-Chatard2012,Koprucki2015}.
One obtains the non-isothermal Scharfetter--Gummel scheme 
\begin{align}
J_{n,K,L} & =M_{n,K,L}k_{B}T_{K,L}g_{n,K,L}\left(n_{L}B\left(X_{n,K,L}\right)-n_{K}B\left(-X_{n,K,L}\right)\right),\label{eq: Scharfetter-Gummel scheme}
\end{align}
(holes analogously) with $X_{n,K,L}=q\left(\Phi_{L}-\Phi_{K}\right)/\left(k_{B}T_{K,L}g_{n,K,L}\right)$
and the Bernoulli function $B\left(x\right)=x/\left(\exp{\left(x\right)}-1\right)$.
The averaged degeneracy factor (consistent with the thermodynamic
equilibrium \cite{Bessemoulin-Chatard2012,Koprucki2015}) and the
logarithmic mean temperature read 
\begin{align}
g_{n,K,L} & =\frac{\eta_{n,L}-\eta_{n,K}}{\log{\left(\mathscr{F}\left(\eta_{n,L}\right)/\mathscr{F}\left(\eta_{n,K}\right)\right)}}, & T_{K,L} & =\Lambda\left(T_{L},T_{K}\right)=\frac{T_{L}-T_{K}}{\log{\left(T_{L}/T_{K}\right)}}.\label{eq: gnKL and TKL}
\end{align}
The scheme (\ref{eq: Scharfetter-Gummel scheme}) is a non-isothermal
generalization of the scheme developed in \cite{Bessemoulin-Chatard2012,Koprucki2015}.

\subsection{Structure-preserving properties}

The discrete system (\ref{eq: discrete system})--(\ref{eq: Scharfetter-Gummel scheme})
has several structure-preserving properties that hold without any
smallness assumption. The conservation of charge is immediately guaranteed
by the finite volume discretization \cite{Farrell2017}. Moreover,
the scheme (\ref{eq: Scharfetter-Gummel scheme}) is robust in both
the drift- and diffusion dominated limits, as it interpolates between
the upwind scheme for $X_{n,K,L}\to\pm\infty$ (strong electric field)
and a central finite difference scheme for $X_{n,K,L}=0$ (pure diffusion).
The latter involves a discrete analogue of the nonlinear diffusion
constant (\ref{eq: generalized Einstein relation}) using $g_{n,K,L}$
as in Eq.~(\ref{eq: gnKL and TKL}). For the analysis of further
properties, which address the consistency with thermodynamics, it
is convenient to recast the formula (\ref{eq: Scharfetter-Gummel scheme})
into a discrete analogue of its thermodynamic form (\ref{eq: current densities - thermodynamic form}):
\begin{equation}
J_{n,K,L}=-\sigma_{n,K,L}\left(\varphi_{n,L}-\varphi_{n,K}+P_{n,K,L}\left(T_{L}-T_{K}\right)\right).\label{eq: SG thermodynamic form}
\end{equation}
The edge-averaged discrete conductivity, which is implicitly taken
by the Scharfetter--Gummel discretization, is a ``tilted'' logarithmic
mean $\Lambda$ of the carrier densities 
\begin{equation}
\sigma_{n,K,L}=\frac{qM_{n,K,L}}{\mathrm{sinhc}{\left(\frac{1}{2}X_{n,K,L}\right)}}\Lambda\left(n_{L}\exp{\left(-\frac{1}{2}X_{n,K,L}\right)},n_{K}\exp{\left(+\frac{1}{2}X_{n,K,L}\right)}\right),\label{eq: discrete conductivity}
\end{equation}
with $\mathop{\mathrm{sinhc}}\left(x\right)=\sinh{\left(x\right)}/x$.
The thermopower $P_{n,K,L}$ (required in Eq.~(\ref{eq: discrete Joule heating}))
reads 
\begin{equation}
\begin{aligned}P_{n,K,L}=-\frac{k_{B}}{q}\bigg[ & \log{\left(\frac{N_{c}\left(T_{L}\right)}{N_{c}\left(T_{K}\right)}\right)}\frac{g_{n,K,L}}{\log{\left(T_{L}/T_{K}\right)}}-\frac{1}{k_{B}}\frac{E_{c}\left(T_{L}\right)-E_{c}\left(T_{K}\right)}{T_{L}-T_{K}}\\
 & \qquad\qquad\qquad-\frac{\left(T_{L}-T_{K,L}\right)\eta_{n,L}-\left(T_{K}-T_{K,L}\right)\eta_{n,K}}{T_{L}-T_{K}}\bigg].
\end{aligned}
\label{eq: discrete thermopower}
\end{equation}
The scheme is manifestly consistent with the thermodynamic equilibrium
(no current for $\varphi_{n,K}=\varphi_{n,L}$ and $T_{K}=T_{L}$)
and the limiting cases of either vanishing chemical ($\varphi_{n,K}=\varphi_{n,L}$:
pure Seebeck current) or thermal ($T_{K}=T_{L}$: isothermal drift-diffusion)
driving forces. The discretization guarantees the non-negativity of
the Joule heat term 
\begin{align}
H_{J,K,L} & =\sum_{\lambda\in\left\{ n,p\right\} }\sigma_{\lambda,K,L}\left|\varphi_{\lambda,L}-\varphi_{\lambda,K}+P_{\lambda,K,L}\left(T_{L}-T_{K}\right)\right|^{2}\geq0\label{eq: non-negativity of Joule heating}
\end{align}
(using Eqs.~(\ref{eq: discrete Joule heating}) and (\ref{eq: SG thermodynamic form}))
and subsequently also the consistency with the 2nd law of thermodynamics
\cite{Kantner2020a}. 
In a 1D case study \cite{Kantner2020a}, the scheme (\ref{eq: Scharfetter-Gummel scheme})
was found to be significantly more accurate than the conventional
Scharfetter--Gummel-type discretization approach. Both schemes revealed
quadratic convergence, but the new scheme (\ref{eq: Scharfetter-Gummel scheme})
saved 1--2 refinement steps to reach the same level of accuracy.

\vspace{-3ex}

\begin{figure}[t]
\includegraphics[width=1\textwidth]{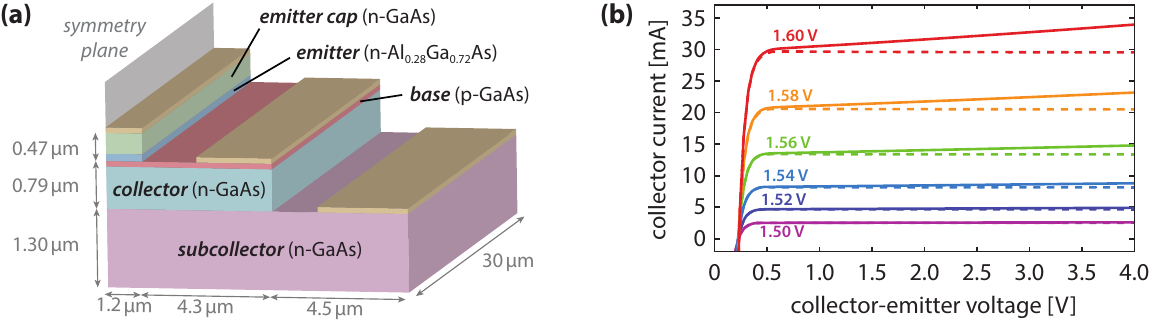}

\caption{\textbf{(a)}~Sketch of the considered GaAs/\,AlGaAs-HBT. Due to
symmetry, only half of the device is simulated. The doping densities
are: $N_{D}^{+}=4\times10^{19}\,\text{cm}^{-3}$ (emitter cap), $N_{D}^{+}=2\times10^{17}\,\text{cm}^{-3}$
(emitter), $N_{A}^{-}=3\times10^{19}\,\text{cm}^{-3}$ (base), $N_{D}^{+}=2\times10^{16}\,\text{cm}^{-3}$
(collector) and $N_{D}^{+}=5\times10^{18}\,\text{cm}^{-3}$ (subcollector).
\textbf{(b)}~Calculated collector current $I_{C}$ as a function
of the collector-emitter voltage $U_{\text{CE}}$ for different base-emitter
voltages $U_{\text{BE}}$ with (solid lines) and without (dashed)
self-heating effects.}

\label{fig: HBT} 
\end{figure}

\section{Numerical simulation of a heterojunction bipolar transistor\label{sec:Numerical-simulation}}

The approach is demonstrated by numerical simulations of the GaAs/\,AlGaAs-based
heterojunction bipolar transistor (HBT) shown in Fig.~\ref{fig: HBT}\,(a).
We assume ideal ohmic contacts with perfect heat sinking ($T_{\text{cont}}=300\,\text{K}$)
and homogeneous Neumann boundary conditions else. The material parameters,
including temperature-dependent models for the band edge energies,
mobilities and the thermal conductivity, are taken from \cite{Palankovski2004}.
The validity of the Kelvin formula for GaAs was studied in \cite{Kantner2020a}.
The calculated current-voltage curves (with and without self-heating
effects) are shown in Fig.~\ref{fig: HBT}\,(b).

\begin{figure}[t]
\includegraphics[width=1\textwidth]{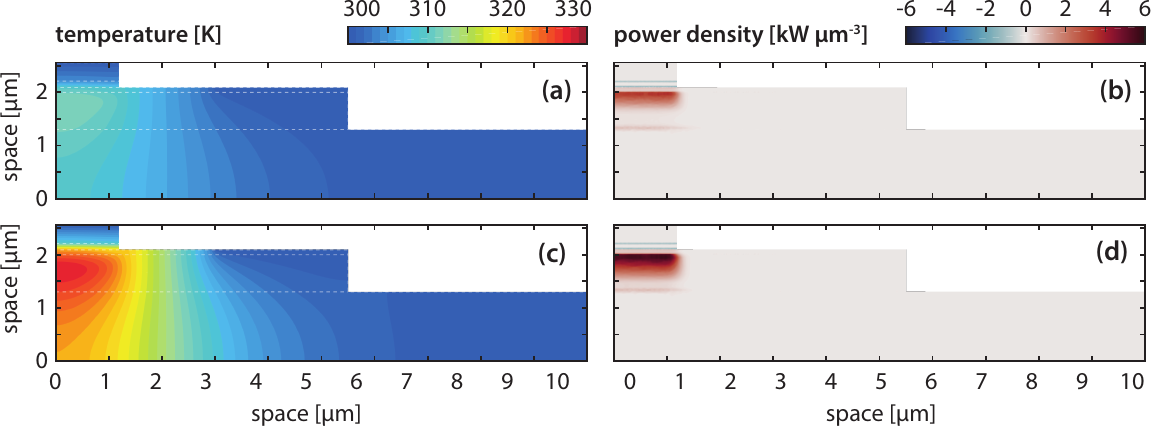}

\caption{Simulated temperature distribution and self-heating power density
$H$ at stationary operation with \textbf{(a,\,b)~}$U_{\text{CE}}=2\,\text{V}$
and \textbf{(c,\,d)}~$U_{\text{CE}}=4\,\text{V}$. The basis-emitter
voltage is $U_{\text{BE}}=1.6\,\text{V}$ in both cases.}
\label{fig: simulation results} 
\end{figure}

The temperature distribution and the heat generation rate are plotted
in Fig.~\ref{fig: simulation results} for different collector-emitter
voltages. The Thomson--Peltier effect is found to cool the AlGaAs/\,GaAs
heterojunctions (emitter/\,emitter cap and emitter/\,base junction,
blue color in Fig.~\ref{fig: simulation results}~(b,\,d)) and
heats up the collector/~subcollector junction. With increasing current
densities (i.e., increasing collector-emitter voltage), the relative
importance of Joule heating increases, until it becomes the dominant
effect. This leads to a strong temperature increase in the collector
region close to the symmetry axis. Recombination processes additionally
heat the base region below the base/\,emitter junction, but were
found to be of minor importance in the present study.

\vspace{-3ex}

\section{Conclusions}

The Kelvin formula for the thermopower yields a remarkably simple
form of the non-isothermal drift-diffusion system. The specific form
of the current density expressions, which contain the thermal driving
forces only implicitly, allow for a non-isothermal generalization
of the Scharfetter--Gummel scheme for Fermi--Dirac statistics that
was previously presented in \cite{Bessemoulin-Chatard2012,Koprucki2015}.
The resulting finite volume scheme preserves fundamental thermodynamic
properties and relations on the discrete level.

\appendix

\section*{Appendix: Derivation of the heat equation \label{sec: appendix}}

In the following, the heat equation \eqref{eq: heat equation} will
be derived from an integral form of the total energy balance equation.
The total energy is obtained from the free energy and the definition
of the entropy (see Sect.~\ref{sec:Kelvin-formula}) as 
\[
\mathcal{E}\left(n,p,T\right)=\mathcal{F}\left(n,p,T\right)+T\mathcal{S}\left(n,p,T\right)=\mathcal{F}\left(n,p,T\right)-T\partial_{T}\mathcal{F}\left(n,p,T\right).
\]
Using the free energy functional (\ref{eq: free energy}), one obtains
\begin{equation}
\begin{aligned}\mathcal{E}\left(n,p,T\right) & =\int_{\Omega}\mathrm{d}V\,\bigg(\frac{TN_{c}^{\prime}\left(T\right)}{N_{c}\left(T\right)}k_{B}TN_{c}\left(T\right)\mathscr{G}\bigg(\mathscr{F}^{-1}\bigg(\frac{n}{N_{c}}\bigg)\bigg)+\left(E_{c}(T)-TE_{c}^{\prime}(T)\right)n\\
 & \hphantom{=\int_{\Omega}\mathrm{d}V}+\frac{TN_{v}^{\prime}\left(T\right)}{N_{v}\left(T\right)}k_{B}TN_{v}\left(T\right)\mathscr{G}\bigg(\mathscr{F}^{-1}\bigg(\frac{p}{N_{v}}\bigg)\bigg)-\left(E_{v}(T)-TE_{v}^{\prime}(T)\right)p\bigg)\\
 & \hphantom{=}+\frac{1}{2}\int_{\Omega}\mathrm{d}V\int_{\Omega}\mathrm{d}V^{\prime}\,G\left(\mathbf{r},\mathbf{r}^{\prime}\right)\rho\left(\mathbf{r}\right)\rho\left(\mathbf{r}^{\prime}\right)+\int_{\Omega}\mathrm{d}V\,\electrostaticPotential_{\text{ext}}\rho+\int_{\Omega}\mathrm{d}V\,u_{L}\left(T\right),
\end{aligned}
\label{eq: total energy functional}
\end{equation}
where $u_{L}=f_{L}\left(T\right)-T\partial_{T}f_{L}\left(T\right)$
is the energy density of the lattice phonons.

The (volumetric) heat capacity of the system is defined as the variational
derivative of the total energy (\ref{eq: total energy functional})
with respect to the temperature\begin{subequations}\label{eq: derivatives of the energy functional}
\begin{equation}
\mathrm{D}_{T}\mathcal{E}\left(n,p,T\right)=c_{V}.\label{eq: derivatives of the energy functional - T}
\end{equation}
Moreover, one obtains 
\begin{align*}
\mathrm{D}_{n}\mathcal{E}\left(n,p,T\right) & =-q\varphi_{n}+k_{B}T\left(\frac{TN_{c}^{\prime}\left(T\right)}{N_{c}\left(T\right)}g\left(\frac{n}{N_{c}}\right)-\mathscr{F}^{-1}\left(\frac{n}{N_{c}}\right)-\frac{1}{k_{B}}E_{c}^{\prime}\left(T\right)\right),\\
\mathrm{D}_{p}\mathcal{E}\left(n,p,T\right) & =+q\varphi_{p}+k_{B}T\left(\frac{TN_{v}^{\prime}\left(T\right)}{N_{v}\left(T\right)}g\left(\frac{p}{N_{v}}\right)-\mathscr{F}^{-1}\left(\frac{p}{N_{v}}\right)+\frac{1}{k_{B}}E_{v}^{\prime}\left(T\right)\right),
\end{align*}
which, assuming the Kelvin formula for the thermopowers (\ref{eq: Seebeck coefficients - explicit}),
can be written as 
\begin{align}
\mathrm{D}_{n}\mathcal{E}\left(n,p,T\right) & =-q\left(\varphi_{n}+TP_{n}\right),\label{eq: derivatives of the energy functional - n}\\
\mathrm{D}_{p}\mathcal{E}\left(n,p,T\right) & =+q\left(\varphi_{p}+TP_{p}\right).\label{eq: derivatives of the energy functional - p}
\end{align}
\end{subequations}The total time derivative of the energy functional
(\ref{eq: total energy functional}) reads 
\begin{align*}
\frac{\mathrm{d}}{\mathrm{d}t}\mathcal{E}\left(n,p,T\right) & =\int_{\Omega}\mathrm{d}V\,\left(\mathrm{D}_{T}\mathcal{E}\left(n,p,T\right)\frac{\partial T}{\partial t}+\mathrm{D}_{n}\mathcal{E}\left(n,p,T\right)\frac{\partial n}{\partial t}+\mathrm{D}_{p}\mathcal{E}\left(n,p,T\right)\frac{\partial p}{\partial t}\right)\\
 & =\int_{\Omega}\mathrm{d}V\,\bigg(c_{V}\frac{\partial T}{\partial t}-q\left(\varphi_{p}+TP_{p}-\varphi_{n}-TP_{n}\right)R\\
 & \hphantom{=\int_{\Omega}\mathrm{d}V\,\bigg(}+\mathbf{j}_{n}\cdot\nabla\left(\varphi_{n}+TP_{n}\right)+\mathbf{j}_{p}\cdot\nabla\left(\varphi_{p}+TP_{p}\right)\bigg)\\
 & \phantom{=}-\oint_{\partial\Omega}\mathrm{d}\mathbf{A}\cdot\big(\left(\varphi_{n}+TP_{n}\right)\mathbf{j}_{n}+\left(\varphi_{p}+TP_{p}\right)\mathbf{j}_{p}\big),
\end{align*}
where we used Eq.~(\ref{eq: derivatives of the energy functional})
and the continuity equations (\ref{eq: electron continuity equation})--(\ref{eq: hole continuity equation}).

The energy dissipated from the system is given by the heat and electrical
energy fluxes leaving the domain through the boundary 
\begin{align}
\frac{\mathrm{d}}{\mathrm{d}t}\mathcal{E}\left(n,p,T\right) & =-\oint_{\partial\Omega}\mathrm{d}\mathbf{A}\cdot\mathbf{j}_{Q}-\int_{\Gamma_{D}}\mathrm{d}\mathbf{A}\cdot\left(\varphi_{n}\mathbf{j}_{n}+\varphi_{p}\mathbf{j}_{p}\right),\label{eq: energy balance equation}
\end{align}
where the heat flux density is known as $\mathbf{j}_{Q}=-\kappa\nabla T+TP_{n}\mathbf{j}_{n}+TP_{p}\mathbf{j}_{p}$
\cite{Kantner2020a}. Here, $\Gamma_{D}\subset\partial\Omega$ denotes
the electrical contacts. On the remaining part of the boundary $\Gamma=\partial\Omega\backslash\Gamma_{D}$,
we assume no-flux boundary conditions $\mathbf{n}\cdot\mathbf{j}_{n/p}=0$,
as the charge carriers can not leave the domain there. Finally, using
the divergence theorem, we obtain the heat transport equation as a
local form of the energy balance equation \eqref{eq: energy balance equation}
\begin{align*}
c_{V}\frac{\partial T}{\partial t}-\nabla\cdot\kappa\nabla T & =-\mathbf{j}_{n}\cdot\nabla\left(\varphi_{n}+TP_{n}\right)-\mathbf{j}_{p}\cdot\nabla\left(\varphi_{p}+TP_{p}\right)\\
 & \phantom{=}\;\,-q\left(\varphi_{p}+TP_{p}-\varphi_{n}-TP_{n}\right)R,
\end{align*}
where the right hand side coincides with the heat generation rate
as given in Eq.~(\ref{eq: heat source term}).

Note that the appealing form (\ref{eq: heat source term}) of the
heat generation rate is a consequence from using the Kelvin formula
for the thermopowers in Eqs.~(\ref{eq: derivatives of the energy functional - n})--(\ref{eq: derivatives of the energy functional - p}).
For different models, additional (transient) terms will occur in the
heat generation rate \cite{Kantner2020a}.
\begin{acknowledgement}
This work was funded by the German Research Foundation (DFG) under
Germany's Excellence Strategy -- EXC2046: \textsc{Math}+ (Berlin
Mathematics Research Center). 
\end{acknowledgement}


\end{document}